\documentclass[useAMS,usenatbib]{mn2e} \input{epsf}

\newif\ifAMStwofonts
\AMStwofontstrue

\usepackage{longtable}
\usepackage{lscape}
\usepackage{float}

\def\lesssim{\mathrel{\hbox{\rlap{\hbox{\lower4pt\hbox{$\sim$}}}\hbox{$<$}}}}
\def\gtrsim{\mathrel{\hbox{\rlap{\hbox{\lower4pt\hbox{$\sim$}}}\hbox{$>$}}}}

\def\l_lsun{$\log{L/\rm L_{\odot}}$~}
\def\masa_msun{$M/ \rm M_{\odot}$~}

\def\m_mstar{$M/M_{*}$~}

\title[PSR~J1417-4402 behaviour]{Does PSR~J1417-4402 system
behave as a ``Redback''?}

\author[M. A. De Vito, J. E. Horvath \& O. G. Benvenuto]
{M. A. De Vito$^{1,2}$\thanks{Member of  the Carrera del Investigador
Cient\'{\i}fico, Consejo Nacional de Investigaciones Cient\'\i ficas y
T\'ecnicas (CONICET). Email: adevito@fcaglp.unlp.edu.ar},
J. E. Horvath$^{3}$\thanks{Email: foton@astro.iag.usp.br},
O. G. Benvenuto$^{1,2}$\thanks{Member of  the Carrera del Investigador
Cient\'{\i}fico, Comisi\'on de  Investigaciones Cient\'{\i}ficas de la Provincia
de Buenos Aires (CIC). Email: obenvenu@fcaglp.unlp.edu.ar}\\
$^{1}$ Instituto de Astrof\'{\i}sica de La Plata, IALP, CCT-CONICET-UNLP,
Argentina\\
$^{2}$ Facultad de Ciencias Astron\'omicas y Geof\'{\i}sicas, Universidad
Nacional de La Plata (UNLP),\\ Paseo del Bosque S/N, B1900FWA, La Plata,
Argentina\\
$^{3}$ Instituto de Astronom\'{\i}a, Geof\'{\i}sica e Ci\^encias
Atmosf\'ericas, Universidade de S\~ao Paulo,\\ R. do Mat\~ao 1226 (05508-090),
Cidade Universit\'aria, S\~ao Paulo SP, Brazil}

\begin{document}

\date{December 19}

\pagerange{\pageref{firstpage}--\pageref{lastpage}} \pubyear{2016}

\maketitle

\label{firstpage}

\begin{abstract}

We study the present evolutionary status of the binary system containing the
2.66~ms pulsar PSR~J1417-4402 in a 5.4~day orbit. This is the pulsar in the
original source 3FGL~J1417.5-4402, that has undergone a transition from  X-ray
state to a pulsar state, just like some redbacks did. The system has many
characteristics similar to redback pulsars family, but is on a much wider orbit.
We show that close binary evolution including irradiation feedback driven by the
luminosity due to accretion onto the neutron star component of the pair, and
evaporation due to pulsar emission, is able to account for the masses of the
components and the photometric data of the donor star.  The tracks leading to
the present  PSR~J1417-4402 are degenerate within a range of parameters,
suggesting that  the {\it same} physics invoked to explain the redback/black
widows groups leads to  the formation of much wider orbit systems, outside the
redback region  limits.

\end{abstract}


\section{Introduction} \label{sec:intro}

Millisecond pulsars (MSPs) are thought to be the final product of close binary
systems (CBSs) evolution, in which a donor star transfers mass and angular
momentum to its companion, an old neutron star (NS).  The first  member of  a
new class of intriguing objects was discovered by \citet{1988Natur.333..237F}.
Since the data was interpreted as a fast NS ablating its companion, as a result
of a former spin-up by accretion, the system was dubbed a ``black widow''(BW).
Since then on, this family of objects has grown very fast, and now they are part
of a group usually refered as  ``spiders'' \footnote{We shall use the
denominations "CBSs" and "spiders" interchangeably in this work, in spite that
some details of these names are not exactly the same. Since our work deals with
intermittent systems this subtelty is not important.}.  Isolated MSPs could
result from the final stage of these systems in which evaporation of the
companion is complete. Later on, a group of accreting systems with comparable
value of its orbital period ($P_{\rm orb}$), but much higher donor masses was
identified, and more recently its number increased considerably due to Fermi/LAT
data. A type of Australian spider (``redback'', RB), the cousin to the North
American black widow spider, are used to name this group. Both groups have
$P_{\rm orb}$ shorter than 1 day, although BW and RB companions have $M_2
\lesssim 0.05\ M_{\odot}$ and $0.1\ M_{\odot} \lesssim M_2 \lesssim 0.7\
M_{\odot}$ (with $M_2$ the mass of the donor star) respectively. A whole
discussion of these systems can be found in \citet{2013IAUS..291..127R}.

From an evolutionary point of view it is relevant to understand the  processes
that give rise to the formation of RB and BW systems. Shortly after the
discovery of the first member of the group, PSR~1957+20, BWs have been
considered as resulting from the evaporation of the donor star driven by pulsar
emission.  Regarding RBs, at present there are different proposals for their
formation. \cite{2013ApJ...775...27C} considered CBS evolution with strong
evaporation, assuming this to be possible because of geometrical effects. They
presented a scenario in which BWs and RBs are consequence of different initial
conditions. \cite{2014ApJ...786L...7B} (BDVH14) considered CBS evolution
including evaporation and irradiation feedback {\it ab initio} (see below for a
description). These phenomena lead to the occurrence of cyclic mass transfer
\citep{2004A&A...423..281B} when the system starts with $P_{\rm orb}$ and $M_2$
in a range typical of RBs progenitors. They concluded that some RBs evolve to
become BWs, while others should form MSP-helium white dwarf (He-WD) pairs.
\cite{2015MNRAS.446.2540S} suggested that RBs are formed by the accretion
induced collapse of oxygen/neon/magnesium WDs. The just-formed NS becomes a
radio pulsar that irradiates and makes the donor star to undergo evaporation.
They assumed that the turn-on of pulsar emission prevents further NS accretion
and claim that in this way it is possible to account for the whole family of
observed RBs.

In the recycled model of pulsars, the onset of accretion causes the NS to spin
up to a hundreds of Hz and the NS mass to grow (see, e. g.,
\citealt{1982Natur.300..728A}).  Standard CBSs evolution predicts a
long-standing (order $10^9$~years) mass transfer episode in which the donor star
loses about 70\% of its mass. During this stage, the system is seen as an X-ray
source, and the pulsar remains hidden. The systems formed by a NS and a low-mass
donor star with a mass $1.0\ M_{\odot} \lesssim M_{2} \lesssim 3.5\ M_{\odot}$,
are known as Low-Mass X-Ray Binaries (LMXBs). After this long mass transfer
episode, a few short-lived  mass transfer events due to hydrogen thermonuclear
flashes may occur. Very little amount of mass is exchanged in these episodes.
The final state of the system is largely determined by the first mass transfer
episode (see, e.g., \citealt{2011ApJ...732...70L} and references therein).

A novel ingredient in the former LMXBs which should be important for relatively
short orbital periods is the so-called irradiation feedback
\citep{1997A&AS..123..273H}. Here, irradiation feedback refers to the heating of
the donor star by the luminosity produced by accretion onto the NS. It should
affect the evolution of the companion and the whole spider system. Including
this effect in evolutionary calculations leads to the occurrence of cyclic mass
transfer episodes as described by \citet{2004A&A...423..281B} instead of the
single long mass transfer episode cited above. Based on those calculations, we
have argued for the existence of an evolutionary connection between BWs and RBs
(see Figure~3 in BDVH14). Also, in this context evaporation means the process of
removal of matter from the donor star due to pulsar emission.

In this paper we shall study the evolution of binary systems that lead to the
formation of RBs in the theoretical framework provided by the models presented
in BDVH14. More specifically, we are interested in the evolutionary state of the
intriguing binary system containing PSR~J1417-4402. The paper is organized as
follow: Section~\ref{sec:summary} summarizes a few recent observations, in
Section~\ref{sec:code} we present the main  characteristics of our code, in
Section~\ref{sec:calcu} we describe the calculations presented in this paper,
and in Section~\ref{sec:conclu} we present our main conclusions.

\section{Summary of recent observations}\label{sec:summary}

Three specific systems have been found giving  support to the recycled scenario
accepted to explain the millisecond pulsar  formation. They behave as RBs
switching between an accretion-disk state,with no radio pulsations (the LMXB
state), and a rotation-powered state  featuring radio pulsations (the PSR
state). These are know as transitional millisecond pulsars (tMSPs).

PSR~J1023+0038, ``the missing link'', is a MSP with $P_{\rm spin} = 1.688$~ms
(where $P_{\rm spin}$ is the spin period of the NS). The orbital period of the
binary system is  $P_{\rm orb} = 0.198$~d. It showed a typical behavior of an
accretion state from May 2000 to December 2001, with clear evidence for the
presence of an accretion disk. From 2002 it was in a quiescent  state
\citep{2009Sci...324.1411A}. There is also evidence that its binary  companion,
with an estimated mass of $0.24~M_{\odot}$ \citep{2012ApJ...756L..25D}, was
filling its Roche lobe. Ten years later,  \citet{2013ATel.5513....1S} reported a
change in its state; and did not detect the pulsar in their radio observations.
\citet{2013ATel.5514....1H} reported the characteristic double-peaked $H\alpha$
line, indicative of the presence of an  accretion disk. The system is presently
again in the LMXB state \citep{2014ApJ...781L...3P}.

PSR~J1824-2452I is another MSP in a spider system, located in the globular
cluster M28. Its companion has a minimum mass of $M_{\rm min} = 0.174 \pm
0.003~M_{\odot}$ (evaluated for a neutron star mass of $1.35~M_{\odot}$ and an
inclination of the system of $90^o$, see \citealt{2013Natur.501..517P}). The
orbital period of the system is $P_{\rm orb} = 0.4594$~d, and the pulsar spin
$P_{\rm spin} = 3.93$~ms. Analysis of the records for this system reveals the
short timescale of the switching mechanism between rotation  and accretion
powered states (PSR and LMXB states respectively). These changes occur in only
few days to months, a lapse far smaller than the characteristic $\sim \, Gyr$
scale for the donor star evolution \citep{2013Natur.501..517P}.

The third member of this transitional group is XSS~J12270-4859, initially
classified as a LMXB with $P_{\rm orb} = 0.2875$~d. \citet{2014MNRAS.441.1825B}
presented radio, optical and X-ray observations showing that XSS J12270-4859 is
a LMXB  that shifted to a radio MSP state between November 14  and December 21
of 2012. \citet{2015MNRAS.454.2190D} presented an analysis of X-ray, UV and
optical/near-IR photometric data obtained after the transition to a
rotation-powered radio PSR state. Their analysis of the optical light curves
give a mass ratio of $0.11 \lesssim q \lesssim 0.26$, whereby the mass of the
donor star is found to be  $0.15\ M_{\odot} \lesssim M_2 \lesssim 0.36\
M_{\odot}$, considering a NS companion with $M_{\rm NS}= 1.4\ M_{\odot}$.

These three transitional systems clearly belong to the RB family.

On the characteristic observational plane used for studying the kind of CBSs
related with recycled pulsars, namely the $M_{2}-P_{\rm orb}$, RBs are located
in a specific region, delimited by the orbital period value of the system and
the estimated mass of the companion. However, a system that is {\it outside} the
RB zone has been recently seen to undergo a transition from a LMXB to a PSR
state, which is quite unexpected.


\citet{2010ApJS..188..405A} discovered a $\gamma$-ray source, detected by the
Large Area Telescope (LAT) on the {\it Fermi Gamma-ray Space Telescope} and
first cataloged as 3FGL~J1417.5-4402. \citet{2013MNRAS.432.1294P} performed
radio continuum observations with the Australian Telescope Compact Array, and
reported no radio emission detected from the source (2012 September 20). In 2013
February 4, it was observed with the 6 km configuration and, again, no radio
emission was detected. \citet{2015ApJ...804L..12S} confirmed the nature of the
optical counterpart, whose spectrum corresponds to a late G or early K star.
Assuming the donor is filling its Roche lobe, they conclude that the object is a
giant star. They further derived an orbital period of $P_{\rm orb} = 5.37385 \pm
0.00035$~d and an eccentricity of $e = 0.01 \pm 0.01$; the mass ratio for the
system is $q = M_{2} / M_{\rm NS} = 0.179 \pm 0.010$. The estimated masses for
the components are $M_{\rm NS} = 1.97 \pm 0.15~M_{\odot}$ and $M_2 = 0.35 \pm
0.04~M_{\odot}$. \citet{2016ApJ...820....6C} confirmed the end of the LMXB stage
when they found a $2.66$~ms pulsar (PSR~J1417-4402) around the $5.4$~day optical
variable discovered by \citet{2015ApJ...804L..12S} as the companion star in  the
high-energy  gamma-ray source 3FGL~J1417.5-4402.

\citet{2016ApJ...820....6C} attribute to eclipses the intervals in which the
pulsar is not detected in their observations. Besides, they deduce that the
delay in the pulse arrival, on at least two occations, is due to an ephemeral
clump of ionized material.

The intrinsic temperature of the unheated companion star of PSR~J1417-4402
deduced from the optical spectrum of is $T_{\rm eff} = 5000$~K
\citep{2015ApJ...804L..12S}. Although some heating of the companion star
photosphere by the pulsar wind should exist, the optical light curves show no
clear evidence of this process, at least at a level of $0.1-0.2$ magnitudes.
This may occur because the intrinsic luminosity of the giant/sub-giant component
dominates the light curve.  \citet{2016ApJ...820....6C} assume an upper limit
for the average temperature for the side facing the pulsar of $T_{\rm h} =
5200$~K. Because of these observational facts, we shall consider that the
temperature of the model should be in the interval $T_{\rm eff} = 5000 -
5200$~K. 

The distance to PSR~J1417-4402 was estimated by \citet{2015ApJ...804L..12S},
$d_S = 4.4$~kpc, based on the magnitude and optical spectrum of the companion
and considering that it fill its Roche lobe.  \citet{2016ApJ...820....6C}
estimate the distance from the radio pulsar dispersion measure (DM), using the
distribution model of \citet{2002astro.ph..7156C}, obtaining $d_C =  1.6$~kpc,
and consider that the best DM-based distance estimation. But it is important to
remark that PSR~J1417-4402 has a high Galactic latitude, and the determination
of DM could have uncertainties. Then $d_S$ can not be rouled out. Recently, a
new model for the large-scale distribution of free electrons in the Galaxy was
presented by \citet{2017ApJ...835...29Y} (know as YMW16 electron density model).
The distance obtained based in this model is $2.16$~kpc
\citep{2017MNRAS.468.3289Y}.  

The typical H$\alpha$ emission line with double-peaked morphology, indicative of
the presence of an accretion disk, has been positively detected by
\citet{2015ApJ...804L..12S}. These authors claim that the X-ray luminosity at
$d_S$ is consistent with those of transitional MSPs in their disk states.

After a detailed analysis, \citet{2016ApJ...820....6C} suggest that
PSR~J1417-4402 is in the radio ejection regime (see Section~\ref{sec:code}), and
this scenario is favored by the orbital period value of this system.  However,
\citet{2015ApJ...807...62A} detected coherent X-ray pulsations from the tMSP
PSR~J1023+0038 in its LMXB sate, interpreted as inflowing material heating the
magnetic polar caps of the NS. This phenomena was seen in the other two know
tMSPs, XSS~J12270-4859 and IGR~J18245-2452, as well as the tMSP candidate
1RXS~J154439.4-112820. This highlights the complexity of the accretion scenario
in this kind of systems (see \citet{2016ApJ...830..122J}).

In a previous paper (BDVH14) we have shown that RBs can experience transitions
between LMXB and PSR states and vice versa. There is no theoretical reason to
expect the occurrence of this behavior only for the RBs group (see Fig.~3 in
BDVH14). Indeed, the zone of cyclic mass transfer episodes in the plane $M_2 -
P_{\rm orb}$ covers a wide range of companion masses and orbital periods. Thus,
systems like PSR~J1417-4402 are naturally expected in the framework of our
models.

With this perspective, we shall now attempt to characterize a possible
progenitor range for the PSR~J1417-4402 system, based on CBS calculations
including irradiation feedback. In other words, we aim to clarify the
evolutionary limits of the transitional systems and, as we shall see,  this
leads us to believe that  should be some systems driven by the same evolution
physics which lay outside the present bounds of the RBs group.

\section{The Close Binary Evolutionary Code} \label{sec:code}

We computed CBS evolution using our code \citep{2003MNRAS.342...50B}, including
irradiation feedback and evaporation. Whenever the pair is detached the code is
fairly standard. When mass transfer conditions are attained it solves the
structure of the donor, mass transfer rate and $P_{\rm orb}$ implicitly. As in
previous works, we assumed that the NS accretes a fixed fraction
$\mathrm{\beta}$ of the transferred matter up to the Eddington critical rate
$\mathrm{\dot{M}_{Edd}\approx 2 \times 10^{-8} M_{\odot}/y}$. Very low accretion
rates ($\mathrm{\lesssim 1.3 \times 10^{-11} M_{\odot}/y}$) are prevented by the
propeller mechanism \citep{2008AIPC.1068...87R}. Nevertheless, there are
observations that show coherent X-ray pulsations at very low luminosities in 
PSR~J1023+0038, that can be asociated to very low mass accretion onto the NS
(see \citealt{2015ApJ...807...62A}). Irradiation feedback has been included
following \citet{1997A&AS..123..273H}. For a complete description see
\citet{2015ApJ...798...44B}. Our models also include evaporation due to pulsar
emission. From an evolutionary point of view the latter represents a minor
perturbation for the donor star at the stages studied in this paper. 

It has  been also been suggested (\citealt{1989ApJ...336..507R};
\citealt{2001ApJ...560L..71B}) that the pressure due to radio pulsar 
emission may inhibit material lost by the donor star to be accreted by the NS.
This phenomenon, usually called {\it radio ejection} may be relevant for  
spider systems. If radio ejection operates, it may inhibit the occurrence of
irradiation feedback. Also, as we have commented above, in some cases
observations indicate the presence of an accretion disk surrounding the NSs.
\citet{2012A&A...537A.104V} studied the evolution of ultra-compact LMXBs,
considering an accretion disk, allowing for the occurrence of short time scale
disk instabilities, modulating the long term evolution computed here. Including
radio ejection and/or accretion disks in our models is beyond the scope of this
paper. 
%
\section{Results} \label{sec:calcu}

\begin{figure}
\epsfysize=300pt
\epsfbox{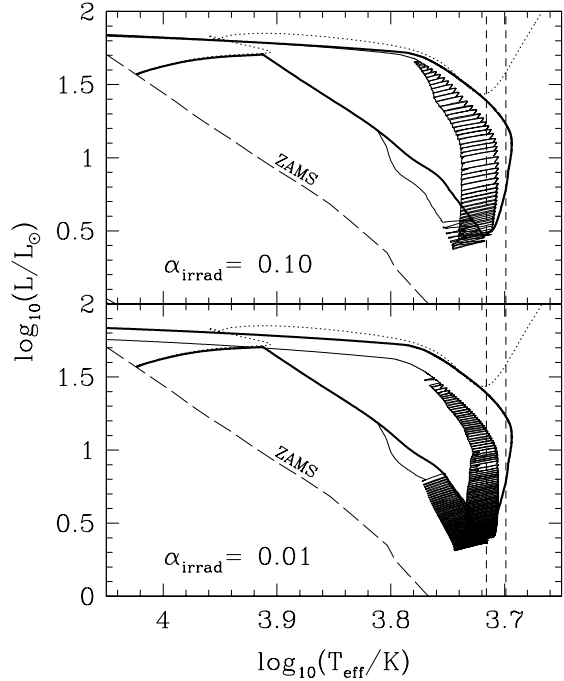}
\caption{The evolution of a solar composition donor star in a spider system with initial
$M_{2}= 2.5\ M_{\odot}$, $M_{\rm NS}= 1.4$, $P_{\rm orb}= 1.4$~days, and $\beta=
0.50$. Upper and lower panels show the cases of $\alpha_{\rm irrad} = 0.10$ and
$0.01$ respectively. In both panels with a thick line we show the non-irradiated
case ($\alpha_{\rm irrad} = 0.00$) and the temperature range corresponding to
this object (discussed in \S~\ref{sec:summary}), with vertical dash lines. Also in
both panels, with dotted line we depict the evolution of an isolated star with
the same mass and chemical composition and, with long dash line, the
position of the zero age main sequence (ZAMS).
\label{Fig:hrs}}
\end{figure}

\begin{figure}
\epsfysize=300pt
\epsfbox{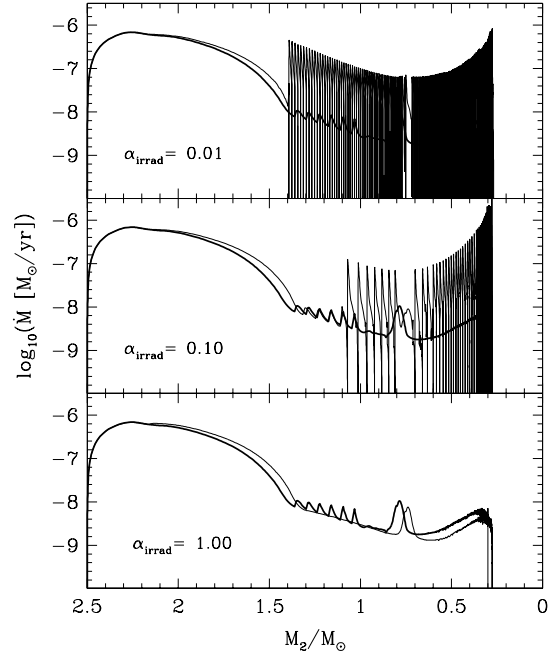}
\caption{The mass transfer rate as a function of the donor mass for the four
model sequences computed for this paper. We show the cases of $\alpha_{\rm
irrad} = 0.01, 0.10$ and $1.00$ in the upper, middle and lower panels,
respectively. In all panels, for comparison, we show with thick solid line the
case of non-irradiated models.
\label{Fig:mdots}}
\end{figure}

\begin{figure}
\epsfysize=300pt
\epsfbox{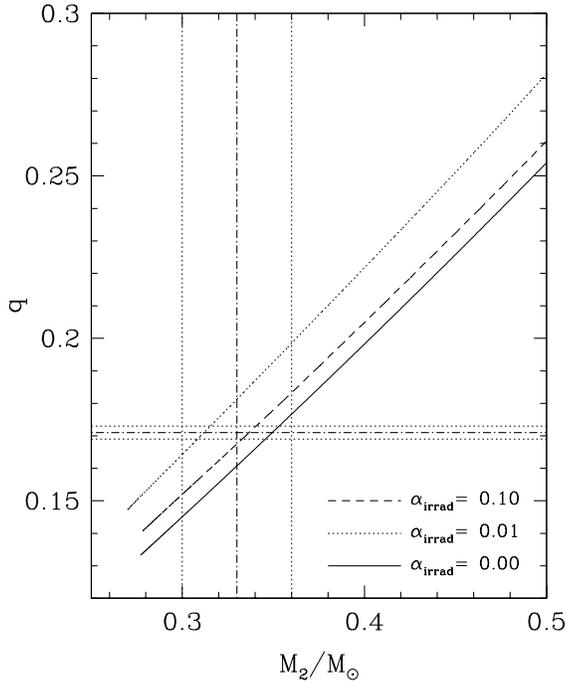}
\caption{The mass ratio $q=M_2/M_{\rm NS}$ as a function of the donor star mass
for the models presented in Fig.~\ref{Fig:hrs}. With vertical and horizontal
dot-dash lines we plot the estimated values of the donor mass and the mass ratio
for the components of this system, respectively; with dot lines, we plot the
errors in each of these estimated values. With full, dot-dash and short-dash
line we depict the evolution of the mass ratio with the mass of the donor star
for the cases of $\alpha_{\rm irrad} = 0.00, 0.01$ and $0.10$, respectively.
\label{Fig:masses}}
\end{figure}

\begin{figure}
\epsfysize=300pt
\epsfbox{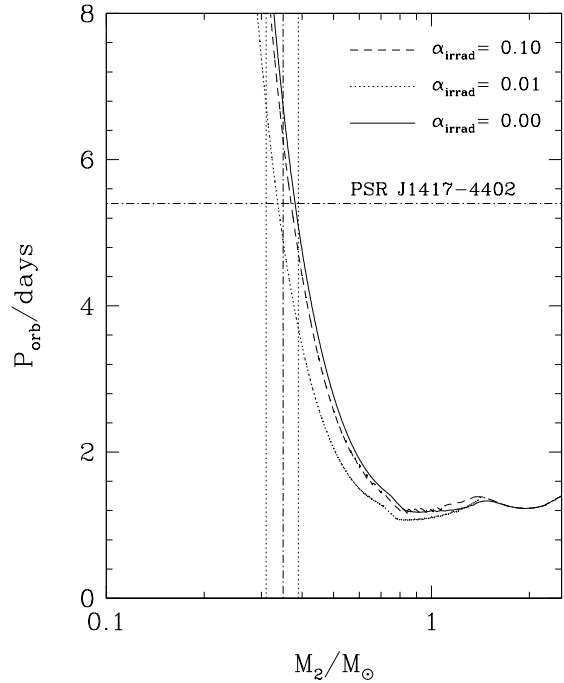}
\caption{The evolution of the orbital period for the models presented in
Fig.~\ref{Fig:hrs} as a function of the mass of the donor star for the same
values of $\alpha_{\rm irrad}$ showed in Figure~\ref{Fig:masses}. The lines have
the same meaning as in that Figure. We indicate with horizontal dash-dot line
the orbital period of the system. Also, with vertical dash-dot line we depict
the mean value of the donor mass whereas dotted lines indicate the 1-$\sigma$
error.
\label{Fig:periodos}}
\end{figure}

\begin{figure}
\epsfysize=300pt
\epsfbox{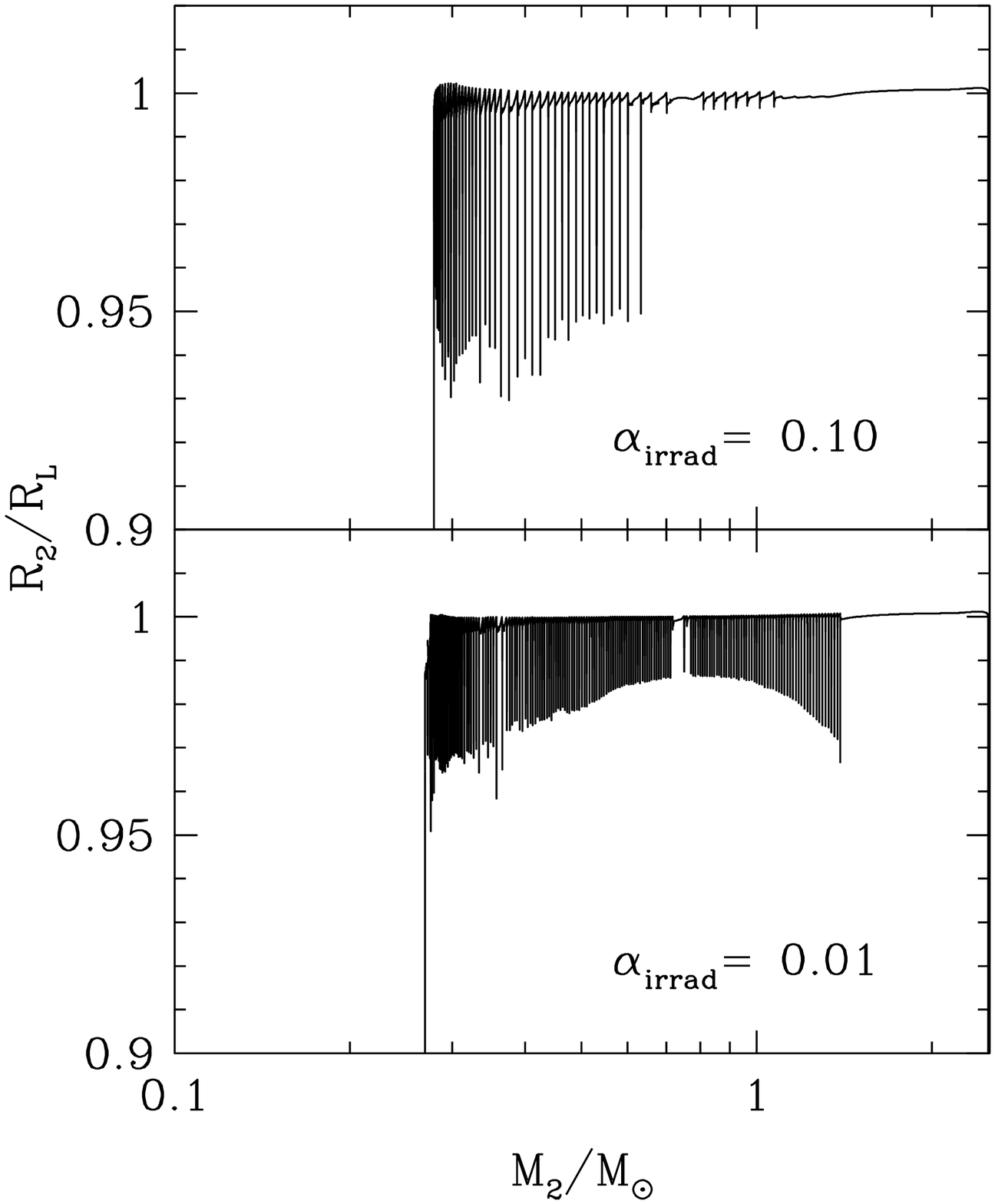}
\caption{The radius of the donor star $R_2$, in units of the Roche lobe radius
$R_{\rm L}$ as a function of the donor star mass for the models presented in
Fig.~\ref{Fig:hrs}. The donor star is in the RLOF or quasi-RLOF states.
\label{Fig:radios}}
\end{figure}

In order to find  a plausible progenitor for the PSR~J1417-4402 system, we
explored the range of initial parameters with possibilities to reproduce the
current state of that system. We based the exploration in our library of
previous calculations \citep{2014ApJ...786L...7B}, where we perfomed a detailed
grid covering initial masses of the donor star between $1.00$ and
$3.50~M_{\odot}$, a $1.40~M_{\odot}$ NS companion, and initial orbital periods
between $0.50$ and $12$~days. We find that a normal, solar composition donor
star with an initial mass value of $M_2 = 2.5\ M_{\odot}$ together with a NS of
$M_{\rm NS} = 1.4\ M_{\odot}$ on a circular orbit with $P_{\rm orb}= 1.4$~days,
and $\beta= 0.50$ was our best candidate. In fact, as we shall show below,
these  initial conditions successfully reproduce the present state of the 
PSR~J1417-4402 system. We considered different strengths for irradiation
feedback controlled by the ``coupling constant'' $\alpha_{\rm irrad}$. This
parameter represents the fraction of the accretion luminosity released by the NS
that effectively participates in the irradiation feedback process. Here we have
considered $\alpha_{\rm irrad} = 0.00, 0.01, 0.10$ and $1.00$. The main results
of this paper are presented in Figs.~\ref{Fig:hrs}-\ref{Fig:radios}.

When considering the case of PSR~J1417-4402 system solely, it would be possible
to argue that the donor star is indeed behaving as predicted by standard
calculations and that the pulsar reveals itself  due to accretion disk
instabilities. However, the population of detected redbacks is rather numerous
(see Fig.~4 of BDVH14), and it seems not plausible to attribute it only to disk
instabilities (see, e.g., \citealt{1998MNRAS.298.1048H}). Irradiated models
depict the main characteristics of redback systems, especially the quasi Roche
lobe overflow state showed by an important fraction of donor stars in this group
of spiders. Besides, the whole spider evolution is nicely accounted for by
irradiated models, like those presented in that paper and in previous
calculations presented by our group (see, e.g., \citealt{2015ApJ...798...44B}).
In the frame of irradiated models, LMXBs and transitional objects should
correspond to systems undergoing mass transfer with the latter undergoing disk
instabilities severe enough to eventually reveal the pulsar. Meanwhile, several
pulsars should be observable due to cessation of mass transfer in each cycle.
Certainly, to be conclusive in this point we would need to perform a population
synthesis analysis, which is beyond the scope of this paper.  We consider that
irradiated models with cyclic mass transfer are more realistic than standard
ones, and this should be valid for the case of PSR~J1417-4402 system too.

In Figure~\ref{Fig:hrs} we show the Hertzsprung-Russell (HR) diagram for the
cases of $\alpha_{\rm irrad} = 0.10$ and $0.01$ in the upper and lower panels
respectively, together with the results corresponding to non-irradiated model.
Based on the range of estimated temperatures for the donor star $log_{10}(T_{\rm
eff}) = 3.699 \lesssim log_{10}T \lesssim log_{10} (T_h) =3.716$
\citep{2016ApJ...820....6C}, the pulsar companion is in an evolutionary state
corresponding to the vertical zone of the right side in the HR diagram.  This
state is completely different from a WD. The He-WD state for the companion of
PSR~J1417-4402 was suggested by \citet{2016ApJ...820....6C} based on the
standard close binary evolution calculation performed by
\citet{2002ApJ...565.1107P}. Moreover, the authors have difficulties trying to
locate the companion star in  this evolutionary state, because it is not clear
if the latter is in a  quasi-Roche lobe overflow (quasi-RLOF) or not (see
\citealt{2015ApJ...798...44B}), if the mass transfer onto the NS has finished or
not, or even if the pulsar is still ablating the companion. Nevertheless our
model naturally fits the observed state of the companion of PSR~J1417-4402.

As we can see from Figure~\ref{Fig:hrs}, irradiated and non-irradiated  models
occupy the same place un the HR diagram. Nevertheless, irradiated models
represent a large improvement in the physical description of these systems,
since they naturally provide, e.g., the quasi-RLOF status observed in many
redback systems.

The evolutionary stage we want to describe for this system is the PSR state.
Observations show a little difference between the pulsar-irradiated and
non-irradiated faces of the donor star.  We could think of attributing this
temperature difference to the homogenization of the residual heat that has
remained in the envelope of the donor star from the LMXB phase, or due to the
heating from the spin-down luminosity of the pulsar in this phase.

Regarding the response of the envelope to changes in the irradiation regime we
should remark that these are quite fast. Let us consider the characteristic
Kelvin-Helmholtz timescale $\tau_{KH}$ given in  \citet{1986A&A...162...71H}

\begin{equation}
\tau_{KH}=\frac{\rho C_p T H_p}{\phi_*}
\end{equation}

where $\rho$ is the density, $C_p$ is the specific heat at constant pressure,
$T$ is the temperature, $H_p=-dr/dln P$ is the pressure scale height, and
$\phi_*$ is the stellar flux. Applying this definition to the case of the our
model for the present status of PSR~J1417-4402, we found results very similar to
those presented in \citet{1986A&A...162...71H}, in particular in the Figures 4a
and 4b of that paper. The outermost layers of the star react on very short
timescales, far shorter than a year. Thus, the difference of 0.017 dex in the
photospheric temperature of the irradiated and non-irradiated portions of the
donor star should not be interpreted as a relic of a previous evolutionary
stages in which the NS acted as a X-ray source but as a consequence of pulsar
spin-down irradiation that is currently received by the star.

Let us estimate the luminosity received by the donor star  to produce a 0.017
dex difference in the irradiated side. If $R$ is the radius of the donor star
and $a$ is the orbital semiaxis, the irradiated and non-irradiated portions of
its surface are $S_{irr}= 2 \pi R^2 (1-R/a)$ and $S_{0}= 2 \pi R^2 (1+R/a)$
respectively. Then, the total luminosity of the donor is $L=2 \pi R^2 \sigma
\big[(1-R/a) T_{irr}^4 + (1+R/a) T_{0}^4 \big]$, where $\sigma$ is the
Stefan-Boltzmann constant and $T_{irr}$ and $T_{0}$ are the temperatures at the
irradiated and non-irradiated portions of the donor photosphere. If we define
$\Delta L$ as the excess of luminosity emitted by the radiated portion of the
donor, $\Delta L= 2 \pi R^2 \sigma (1-R/a)\big[T_{irr}^4 -T_{0}^4 \big]$, and
consider the case of PSR~J1417-4402 where we have $R= 4.8 R_{\odot}$, $R/a=
0.24$, $\log{T_{irr}}= 3.716$, and  $\log{T_{0}}= 3.699$, $\Delta L= 0.8359
L_{\odot}$.

Observed from the NS distance, the donor star represents a solid angle $\Delta
\Omega= 2 \pi \big[1-\sqrt{1-(R/a)^2}]$, that for PSR~J1417-4402 is $\Delta
\Omega= 0.1836$. If for simplicity we consider isotropic emission, it
corresponds to a pulsar luminosity of $L_{psr}= 57.2 L_{\odot}$. If we assume
that this stems from the rotational energy $L_{psr}= - 2 \pi^2 I_{NS} \dot{P}
P^{-3}$ where $P$ is the pulsar period, $\dot{P}$ its time derivative and
$I_{NS}$ is the moment of inertia of the NS. For PSR~J1417-4402 $P= 2.66$~ms, if
we consider a typical value for the moment of inertia of $I_{NS}= 10^{45}\; g\;
cm^2$ we find $\dot{P}= 2.06 \times 10^{-19} s/s$ which is in the range of
values observed for several millisecond pulsars. Considering this infered value
for $\dot{P}$ we obtain a magnetic field strength of  $7.5 \times 10^{8}~G$,
using $B \simeq 3.2 \times 10^{19} (P[s] \, \dot{P}[s/s])^{1/2}~G$
\citep{1977puls.book.....M}. This infered value is in the range corresponding to
recicled pulsars $\sim 10^{8} - 10^{9}~G$.

Figure~\ref{Fig:mdots} shows the mass transfer rate for the three sequences of
irradiated models considered in this paper, and the non-irradiated model with a
thick line. As we can see from this Figure, the model corresponding to very high
irradiation $\alpha_{\rm irrad} = 1.00$ does not undergo cyclic mass transfer
around of the mass value estimated for the companion of PSR~J1417-4402. Thus, we
rule out  this extreme model as a possible progenitor for this system.

In Figure~\ref{Fig:masses} we show the evolution of the mass ratio $q = M_2 /
M_{\rm NS}$ as a function of the the donor star mass, $M_2$. From this graph we
conclude that our best candidate for the progenitor of the system containing
PSR~J1417-4402 is the corresponding to $\alpha_{\rm irrad} = 0.10$, since it is
able to account for the correct values of $q$ and $M_2$, and therefore that of
$M_{\rm NS}$.

In Figure~\ref{Fig:periodos} we show the evolution of the orbital period as a
function of the donor star mass for the cases of $\alpha_{\rm irrad} = 0.00,
0.01$ and $0.10$. As we can see from this Figure, the progenitor with
$\alpha_{\rm irrad} = 0.10$ also reproduces the appropriate value of the orbital
period at the correct value of the donor star mass.

Finally, we show in Figure~\ref{Fig:radios} the evolution of the ratio of the
radii of the donor star to that of its equivalent Roche lobe as a function of
its mass for $\alpha_{\rm irrad} = 0.10$ and $0.01$. As it stands, the donor
star is in the quasi-RLOF state \citep{2015ApJ...798...44B}.

These results show that models considering irradiation feedback and evaporation
are able to account for the main characteristics of the system containing
PSR~J1417-4402. Nevertheless, we want to remark that this solution is one of the
possible ways to attain the current state of PSR J1417-4402. It represent the
best solution based on our own hypothesis and calculations, but other
simulations with different hypothesis and initial conditions could account for
the evolution and the current state of this system as well.

\section{Discussion and Conclusions} \label{sec:conclu}

In this paper we have studied the possibility that the evolution of CBSs
naturally accounts for the present status of the system containing
PSR~J1417-4402. For this purpose we have searched for systems that evolve to the
masses and orbital period values deduced from observations. We have found that
models with initial parameters $M_{\rm NS}= 1.4\ M_{\odot}$,  $M_{2}= 2.5\
M_{\odot}$, $P_{\rm orb}= 1.4$~days, and $\beta= 0.5$ evolve to get the required
donor ans NS masses at the observed orbital period. For a positive detection  of
the pulsar emission, the system should be detached, a requirement fulfilled by
models with $\alpha_{\rm irrad} = 0.10$, being in the quasi-RLOF state.

Regarding our assumption on the efficiency parameter $\beta$, we should remark
that the value assumed in this paper is usually employed in the literature
(see,  e.g., \citealt{2002ApJ...565.1107P}). However, it is possible (and even
reasonable)  that $\beta$ takes  a different value and/or be variable depending
on the evolutionary state of the spider system. In any case, we should stress
that considering different values of $\beta$ will affect the initial values of
the CBS suitable to account for this system, suggesting a degeneracy of the
parameters leading to the desired state.

It is important to remark that while irradiation feedback is responsible for the
occurrence of cyclic mass transfer episodes, evaporation is a necessary
ingredient of the model, because of the observations of eclipses due to material
present far beyond the Roche lobe of the donor star. In conditions of quasi-RLOF
the donor star should be evaporating. In any case, the standard treatment of
evaporation indicates a mass loss rate which is too low to cause a major effect
on the evolution of the donor star at that stage. Also, as already stated,
material lost from the system due to non-conservative mass transfer in previous
RLOFs should also participate in the eclipses.

PSR~J1417-4402 has an orbital period falling outside the parameter region
usually considered for RBs. Current models indicate that the donor star of this
system should evolve to become a He-WD. From a theoretical point of view and
based in our irradiated models, the system not behave differently from those
``standard RBs''. Indeed, it can be checked that  PSR~J1417-4402 is located in
the cyclic mass transfer region in Figure~3 of BDVH14. Therefore, based on our
models with irradiation and evaporation, the observed behavior can be considered
as expected.

The neighborhood of the position occupied by the companion of PSR~J1417-4402 in
the $M_2 - P_{\rm orb}$ plane is the region populated by CO-WDs companions. But,
the companion of PSR~J1417-4402 is clearly an extended, non-degenerate star (as
the observations have shown, and has been depicted by our models). Then, we
expect a mixed  population of pulsars companions above of the usually considered
redback region. 

We want to thank our referee for his/her report, that lead us to fruitful
discussions that contributed to largely improve our paper.


J.E.H. has been supported by Fapesp (S\~ao Paulo, Brazil) through the grant
2013/26258-4 and CNPq, Brazil funding agencies.


\label{lastpage}


\begin{thebibliography}{99}

\bibitem[\protect\citeauthoryear{Abdo et al.}{2010}]{2010ApJS..188..405A} Abdo
A.~A., et al., 2010, ApJS, 188, 405

\bibitem[\protect\citeauthoryear{Alpar et al.}{1982}]{1982Natur.300..728A}
Alpar M.~A., Cheng A.~F., Ruderman M.~A., Shaham J., 1982, Natur, 300, 728

\bibitem[\protect\citeauthoryear{Archibald et al.}{2009}]{2009Sci...324.1411A}
Archibald A.~M., et al., 2009, Sci, 324, 1411

\bibitem[\protect\citeauthoryear{Archibald et al.}{2015}]{2015ApJ...807...62A}
Archibald A.~M., et al., 2015, ApJ, 807, 62 

\bibitem[\protect\citeauthoryear{Bassa et al.}{2014}]{2014MNRAS.441.1825B}
Bassa C.~G., et al., 2014, MNRAS, 441, 1825

\bibitem[\protect\citeauthoryear{Benvenuto \& De
Vito}{2003}]{2003MNRAS.342...50B} Benvenuto O.~G., De Vito M.~A., 2003, MNRAS,
342, 50

\bibitem[\protect\citeauthoryear{Benvenuto, De Vito, \&
Horvath}{2014}]{2014ApJ...786L...7B} Benvenuto O.~G., De Vito M.~A., Horvath
J.~E., 2014, ApJ, 786, L7 (BDVH14)

\bibitem[\protect\citeauthoryear{Benvenuto, De Vito, \&
Horvath}{2015}]{2015ApJ...798...44B} Benvenuto O.~G., De Vito M.~A., Horvath
J.~E., 2015, ApJ, 798, 44

\bibitem[\protect\citeauthoryear{B{\"u}ning \& Ritter}{2004}]
{2004A&A...423..281B} B{\"u}ning A., Ritter H., 2004, A\&A, 423, 281  647

\bibitem[\protect\citeauthoryear{Burderi et al.}{2001}]{2001ApJ...560L..71B}
Burderi L., et al., 2001, ApJ, 560, L71

\bibitem[\protect\citeauthoryear{Camilo et al.}{2016}]{2016ApJ...820....6C}
Camilo F., et al., 2016, ApJ, 820, 6

\bibitem[\protect\citeauthoryear{Chen et al.}{2013}]{2013ApJ...775...27C} Chen
H.-L., Chen X., Tauris T.~M., Han Z., 2013, ApJ, 775, 27

\bibitem[\protect\citeauthoryear{Cordes \& Lazio}{2002}]{2002astro.ph..7156C}
Cordes J.~M., Lazio T.~J.~W., 2002, astro, arXiv:astro-ph/0207156

\bibitem[\protect\citeauthoryear{Deller et al.}{2012}]{2012ApJ...756L..25D}
Deller A.~T., et al., 2012, ApJ, 756, L25

\bibitem[\protect\citeauthoryear{de Martino et
al.}{2015}]{2015MNRAS.454.2190D} de Martino D., et al., 2015, MNRAS, 454, 2190

\bibitem[\protect\citeauthoryear{Fruchter, Stinebring, \&
Taylor}{1988}]{1988Natur.333..237F} Fruchter A.~S., Stinebring D.~R., Taylor
J.~H., 1988, Natur, 333, 237

\bibitem[Halpern \emph{et al.}(2013)]{2013ATel.5514....1H} Halpern, J.P.,
Gaidos,E., Sheffield, A., Price-Whelan, A.M., and Bogdanov, S.: 2013, The
Astronomer's Telegram, 5514.

\bibitem[\protect\citeauthoryear{Hameury, King, \&
Lasota}{1986}]{1986A&A...162...71H} Hameury J.~M., King A.~R., Lasota
J.~P., 1986, A\&A, 162, 71 

\bibitem[\protect\citeauthoryear{Hameury \&
Ritter}{1997}]{1997A&AS..123..273H} Hameury J.-M., Ritter H., 1997, A\&AS,
123,

\bibitem[\protect\citeauthoryear{Hameury et al.}{1998}]{1998MNRAS.298.1048H}
Hameury J.-M., Menou K., Dubus G., Lasota J.-P., Hure J.-M., 1998, MNRAS, 298,
1048 


\bibitem[\protect\citeauthoryear{Jaodand et al.}{2016}]{2016ApJ...830..122J}
Jaodand A., Archibald A.~M., Hessels J.~W.~T., Bogdanov S., D'Angelo C.~R.,
Patruno A., Bassa C., Deller A.~T., 2016, ApJ, 830, 122 

\bibitem[\protect\citeauthoryear{Lin et al.}{2011}]{2011ApJ...732...70L} Lin
J., Rappaport S., Podsiadlowski P., Nelson L., Paxton B., Todorov P., 2011,
ApJ, 732, 70

\bibitem[\protect\citeauthoryear{Manchester \&
Taylor}{1977}]{1977puls.book.....M} Manchester, R.~N., Taylor, J.~H.\ 1977.\
Pulsars.\ San Francisco: W.~H.~Freeman

\bibitem[\protect\citeauthoryear{Papitto et al.}{2013}]{2013Natur.501..517P}
Papitto A., et al., 2013, Natur, 501, 517

\bibitem[\protect\citeauthoryear{Patruno et al.}{2014}]{2014ApJ...781L...3P}
Patruno A., et al., 2014, ApJ, 781, L3

\bibitem[\protect\citeauthoryear{Petrov et al.}{2013}]{2013MNRAS.432.1294P}
Petrov L., Mahony E.~K., Edwards P.~G., Sadler E.~M., Schinzel F.~K.,
McConnell D., 2013, MNRAS, 432, 1294

\bibitem[\protect\citeauthoryear{Podsiadlowski, Rappaport, \&
Pfahl}{2002}]{2002ApJ...565.1107P} Podsiadlowski P., Rappaport S., Pfahl
E.~D., 2002, ApJ, 565, 1107

\bibitem[\protect\citeauthoryear{Roberts}{2013}]{2013IAUS..291..127R} Roberts
M.~S.~E., 2013, IAUS, 291, 127

\bibitem[\protect\citeauthoryear{Romanova et al.}{2008}]{2008AIPC.1068...87R}
Romanova M.~M., Kulkarni A.~K., Long M., Lovelace R.~V.~E., 2008, AIPC, 1068,
87

\bibitem[\protect\citeauthoryear{Ruderman, Shaham, \&
Tavani}{1989}]{1989ApJ...336..507R} Ruderman M., Shaham J., Tavani M., 1989,
ApJ, 336, 507

\bibitem[\protect\citeauthoryear{Smedley et al.}{2015}]{2015MNRAS.446.2540S}
Smedley S.~L., Tout C.~A., Ferrario L., Wickramasinghe D.~T., 2015, MNRAS, 446, 2540

\bibitem[Stappers \emph{et al.}(2013)]{2013ATel.5513....1S}Stappers, B.W.,
Archibald, A., Bassa, C., Hessels, J., Janssen, G., Kaspi, V., Lyne, A., Patruno,
A., and Hill, A.B.: 2013, {\it The Astronomer's Telegram}, 5513

\bibitem[\protect\citeauthoryear{Strader et al.}{2015}]{2015ApJ...804L..12S}
Strader J., et al., 2015, ApJ, 804, L12


\bibitem[\protect\citeauthoryear{van Haaften et
al.}{2012}]{2012A&A...537A.104V} van Haaften L.~M., Nelemans G., Voss R., Wood
M.~A., Kuijpers J., 2012, A\&A, 537, A104

\bibitem[\protect\citeauthoryear{Yao, Manchester, \&
Wang}{2017a}]{2017ApJ...835...29Y} Yao J.~M., Manchester R.~N., Wang N., 2017,
ApJ, 835, 29 

\bibitem[\protect\citeauthoryear{Yao, Manchester, \&
Wang}{2017b}]{2017MNRAS.468.3289Y} Yao J.~M., Manchester R.~N., Wang N., 2017,
MNRAS, 468, 3289 

\end{thebibliography}
\end{document}